\documentclass[epj]{svjour}
\usepackage{epsfig}

\newcommand{\equ}[1]{(\protect\ref{#1})}

\begin{document}

\title{Reaction-diffusion system with self-organized critical behavior} 
 
\author{Romualdo Pastor-Satorras\inst{1} \and Alessandro
  Vespignani\inst{2}} 

\institute{Dept. de F\'{\i}sica i Enginyeria Nuclear,
  Universitat Polit\`{e}cnica de Catalunya,
  Campus Nord, M\`{o}dul B4,  08034 Barcelona, Spain \and 
  The Abdus Salam International Centre for Theoretical Physics
  (ICTP),  P.O. Box 586, 34100 Trieste,
  Italy}

\date{\today}

\abstract{We describe the construction of a conserved
  reaction-diffusion system that exhibits self-organized critical
  (avalanche-like) behavior under the action of a slow addition of
  particles. The model provides an illustration of the general
  mechanism to generate self-organized criticality in conserving
  systems.  Extensive simulations in $d=2$ and $3$ reveal critical
  exponents compatible with the universality class of the stochastic
  Manna sandpile model.
}

\maketitle

Since the introduction of the Bak, Tang, and Wiesenfeld sandpile model
\cite{btw1}, the concept of self-organized criticality (SOC)
\cite{jensen98} has witnessed a real explosion of activity, covering
both the description of new models and the proposal of several
theoretical approaches, aiming at an understanding of SOC phenomena in
terms of standard statistical mechanical concepts.  At this respect,
it has been shown that SOC in sandpile models is related to the
behavior of absorbing-state phase transitions (APT) with
many absorbing states \cite{dvz98,vdmz00}.  Indeed, this very idea is
underlying a recipe proposed for the construction of SOC models
\cite{brazil}: any cellular automata, defined with {\em conserved}
microscopic rules, and possessing many absorbing states, will display
SOC behavior if {\em slowly driven} with the addition of
energy/particles at a rate $h$ and dissipation at a rate $\epsilon$:
i.e., in the double limit $h\to 0$, $\epsilon\to 0$, with $h/\epsilon
\to 0$ \cite{grinstein95}.
This mechanism is easily seen at work in all standard sandpile models
proposed so far \cite{btw1,zhang89,manna91b}. 

Given that most SOC
systems are defined in terms of sandpile-like models (with the
exception of the forest-fire \cite{drossel92} and extremal
\cite{bak93} models), it becomes all the most interesting to explore
the possibility of applying the recipe of Ref.~\cite{brazil} to models
of a qualitatively different sort.  In this paper we consider a
reaction-diffusion (RD) model showing an APT that conserves the total
number of particles \cite{wij98,pv00}.  This model exhibits a
non-equilibrium phase transition in the same universality class of
fixed energy stochastic sandpiles~\cite{vdmz00,pv00}.  Here, we show
that implementing the slow driving condition, the model reaches a
stationary state with an avalanche-like reaction activity with
critical properties. By measuring usual magnitudes characterizing the
SOC behavior, we compare the model with standard slowly driven
sandpiles. The critical exponents measured confirm the shared
universality class with stochastic sandpiles, and provide a vivid
illustration of the SOC generating mechanism \cite{brazil}.

We focus on the two species RD system \cite{wij98,pv00}, recently
proposed to describe APT coupled to a non-diffusive conserved field
\cite{rossi00}. The RD system is defined by the following set of
reaction steps:
\begin{eqnarray}
  B &\to& A \quad \mbox{\rm with rate $k_1$}, \label{reaction1}\\
  B + A &\to& 2 B  \quad \mbox{\rm with rate $k_2$}.\label{reaction2} 
\end{eqnarray}
In this system, $B$ particles diffuse with diffusion rate $D_B$, and A
particles {\em do not diffuse}, that is, $D_A=0$. From the rate
Eqs.~\equ{reaction1} and~\equ{reaction2}, it is clear that the
dynamics conserves the total number of particles $N=N_A+N_B$, where
$N_i$ is the number of particles $i=A, B$. In this model, the dynamics
is exclusively due to $B$ particles, that we identify as {\em active}
particles. $A$ particles do not diffuse and cannot generate
spontaneously $B$ particles.  More specifically, $A$ particles can
only move via the motion of $B$ particles that later on transform into
$A$ through Eq.~(\ref{reaction2}).  This implies that any
configuration devoid of $B$ particles is an absorbing state in which
the system is trapped forever. The number of these absorbing states is
infinite---in the thermodynamic limit---corresponding to all the
possible redistributions of $N$ particles of type $A$ in the system.
This RD process exhibits a phase transition from an active phase (with
an everlasting activity of B particles) to an absorbing phase (no B
particles) for a critical value $\rho=\rho_c$ of the total particle
density \cite{pv00}.

Here, we define a driven-dissipative version of the RD model by
applying the recipe of Refs. \cite{vdmz00,brazil}.  On hypercubic
lattices of size $L$ with open boundary conditions, each site $i$
stores a number $a_i$ of $A$ particles and $b_i$ of $B$ particles. The
occupation numbers $a_i$ and $b_i$ can have any integer value,
including $a_i=b_i=0$, that is, void sites with no particles. The
model is thus representing the dynamics of {\em bosonic} particles.
The initial configuration is constructed by randomly distributing a
number $N_0$ of $A$ particles in the lattice. The initial occupation
numbers $a_i$ have a poissonian distribution, while $b_i=0, \forall
i$. Any configuration is stable whenever it fulfills this condition,
i.e., in the absence of $B$ particles. The system is driven by adding
one $B$ particle to a randomly chosen site. A state with at least one
$B$ particle is called active. Active states evolve in time according
to the following update rules, that mimic the diffusion and reaction
steps in the RD system: I) Diffusion: on each lattice site, each $B$
particle moves into a randomly chosen nearest neighbor site with
probability $2 d /(2 d +1)$, and stays in the same site with
probability $1 /(2 d +1)$; this results in an effective diffusivity
$D_B=1 /(2 d +1)$.  II) After all sites have been updated for
diffusion, we perform the reactions: a) On each lattice site, each $B$
particle is turned into an $A$ particle with probability $r_1$. b) At
the same time, each $A$ particle becomes a $B$ particle with
probability $1-(1-r_2)^{b_i}$, where ${b_i}$ is the total number of
$B$ particles in that site. This corresponds to the average
probability for an $A$ particle of being involved in the
reaction~(\ref{reaction2}) with any of the $B$ particles present on
the same site.  The probabilities $r_1$ and $r_2$ are related to the
reaction rates $k_1$ and $k_2$ defined in Eqs.~\equ{reaction1} and
\equ{reaction2}.  In general, we have that $r_i(k_i=0)=0$,
$r_i(k_i=\infty)=1$, and $r_i$ is an increasing function of $k_i$.
The analytic expression of $r_i$ as a function of $k_i$ is presumably
quite complex and nontrivial. However, as we will argue later, the
knowledge of the precise relationship between $r_i$ and $k_i$ is not
necessary, since the critical behavior of the model should be
independent of the exact values of the parameters $r_i$ selected.  $B$
particles on boundary sites may choose to diffuse out of the lattice.
In this case, the particle is removed out of the system, contributing
to the dissipation. The system is updated in parallel until there are
no more $B$ particles and it is again in an absorbing state. During
the dynamic evolution, the addition of new $B$ particles is suspended;
this of course corresponds to the slow-driving condition.  The
sequence of updates in the system (from the time we introduce a new
$B$ particle until a stable state is reached) is interpreted as an
avalanche. We characterize avalanches by their size $s$ and their
duration $t$. The size of an avalanche is defined as the number of $B$
particles present in the system at each time step, summed over all the
parallel updates required to reach a new stable state. The duration of
an avalanche is defined as the total number of parallel updates
performed during the avalanche.  In
the slow driving perspective, the existence of a critical stationary
state is easily understood. Particles are added only in the absence of
activity ($\rho<\rho_c$), while dissipation acts only during activity
($\rho>\rho_c$). This implies that $\partial_t\rho$ always drives the
system toward $\rho_c$, that in the thermodynamic limit is the only
possible stationary value of the density \cite{vdmz00,brazil}.

\begin{figure}[t]
  \centerline{\epsfig{file=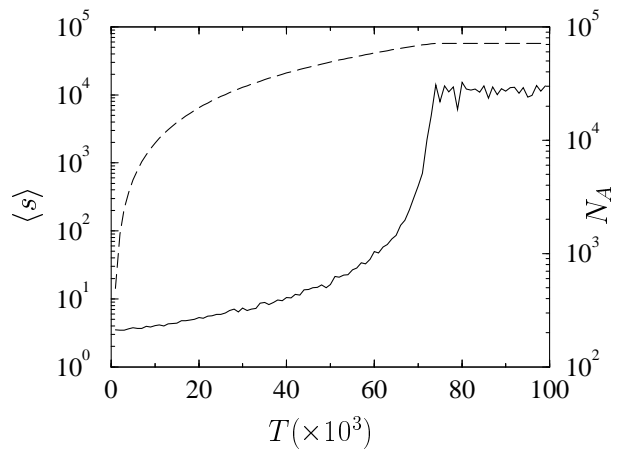, width=8cm}}
  \caption{Number of $A$ particles $N_A$ (dashed  line) and
    average avalanche size $\left< s \right>$ (full line) as a
    function of the number of avalanches $T$ in a slowly driven
    conserved reaction diffusion system in a lattice of size $L=256$.
    The initial state is an empty lattice.}
  \label{fig:New}
\end{figure}

We have performed numerical simulations of this model in dimensions
$2$ and $3$, with system sizes ranging from $L=64$ to $L=1024$ in
$d=2$, and from $L=74$ to $L=280$ in $d=3$. The reaction rates $r_i$
reported here are $r_1=0.3$ and $r_2=0.4$ in $d=2$, and $r_1=0.4$ and
$r_2=0.5$ in $d=3$; the numerical results however, are independent of
the particular values of the reaction rates $r_i$ employed. The
independence of the dynamics from the particular parameters $r_i$
chosen can be easily grasped by the mean-field approach reported in
Ref.~\cite{wij98}, that shows the existence of a single critical
point. More precisely, this can be understood in terms of
renormalization-group arguments: The critical behavior of the model is
ruled by the fixed point toward which the parameters flow under an
appropriate renormalization-group transformation \cite{wij98}. Thus,
the critical exponents are independent of the initial values of the
parameters, and depend only on the value of the unique fixed point. It
is then natural that simulations performed with different $r_i$
parameters will yield the same steady state behavior. This fact is
confirmed by the numerical simulations, which show differences only in
the transient regime.  After the transient initial regime (whose
length depends on $L$, $r_i$, and the initial concentration $N_0$ of
$A$ particles), the system reaches a steady state in which the stable
configurations have a constant average number $\bar{N}_A$ of $A$
particles and avalanches have a constant average size $\left< s
\right>$. As an example, in Fig.~\ref{fig:New}, we plot the number of
$A$ particles and the average avalanche size as a function of the
number of avalanches $T$ in a simulation with system size $L=256$ in
$d=2$, starting from an empty lattice.  In analogy with sandpiles and
SOC phenomena, in the steady state, we compute the probability
distribution of sizes $P(s)$ and times $P(t)$ for the reaction events.
By assuming the usual finite-size scaling form (FSS) \cite{cardy88}
\begin{eqnarray}
  P(s,L) &=& s^{-\tau_s}{\cal F} \left(\frac{s}{L^D}\right),
  \label{eq:fssS}\\
  P(t,L) &=& t^{-\tau_t}{\cal G} \left(\frac{t}{L^z}\right),
  \label{eq:fssT}
\end{eqnarray}
we can define the standard critical exponents, $\tau_s$, $\tau_t$, $D$
(the fractal dimension), and $z$ (the dynamic critical exponent),
which characterize the universality class of the model.  Averages are
performed over at least $5 \times 10^6$ a\-va\-lan\-ches in $d=2$.
The distributions on $d=3$ are considerably noisier; averages have
thus been done in this case over $10^7$ avalanches. In order to check
the plausibility of the FSS hypothesis \equ{eq:fssS} and
\equ{eq:fssT}, and compute the corresponding critical exponents, we
have applied the moment analysis technique, introduced in
Refs.~\cite{men}.  We define the $q$-th moment of the avalanche size
distribution on a lattice of size $L$ as $\left< s^q \right>_L = \int
ds \, s^q \, P(s, L)$. If the FSS hypothesis \equ{eq:fssS} is valid in
the asymptotic limit of large $s$, then the $q$-th moment has the
following dependence on system size:
\begin{equation}
  \left< s^q \right>_L = L^{D(q+1-\tau_s)} \int dy \, y^{(q-\tau_s)}\, 
  {\cal   F}(y) \sim L^{\sigma_s(q)}.
  \label{eq:moments}
\end{equation}
The exponent $\sigma_s(q)=D(q+1-\tau_s)$ is computed as the slope of
the log-log plot of $\left< s^q \right>_L$ as a function of $L$. For
large enough values of $q$ (i.e., away from the region where the
integral in \equ{eq:moments} is dominated by its lower cut-off), one
can compute the fractal dimension $D$ as the slope of $\sigma_s(q)$ as
a function of $q$: $D=\partial \sigma_s(q)/ \partial q$. Once $D$ is
known we can estimate $\tau_s$ using the relation
$\sigma_s(1)=D(2-\tau_s)$. The exponent $\sigma_s(1)$, giving the
scaling of the average avalanche size, can be estimated using a
standard argument in sandpiles: a new injected particle $B$ performs a
diffusing Brownian motion and has to travel, on average, a distance of
order $L^2$ before reaching the boundary. In the stationary state, to
each $B$ particle drop must correspond, on average, a $B$ particle
diffusing out of the system.  This implies that the average avalanche
size corresponds to the average number of diffusion steps needed for a
$B$ particle to reach the boundary; i.e., $\left<s\right>\sim L^2$,
and thus $\sigma_s(1)=2$.  Along the same lines we can obtain the
moments of the avalanche time distribution. In this case, $\langle
t^q\rangle_L\sim L^{\sigma_t(q)}$, with $\partial\sigma_t(q)/\partial
q=z$. Analogous considerations for small $q$ apply also for the time
moment analysis.  Then, the $\tau_t$ exponent can be found using the
scaling relation $z(2-\tau_t)=\sigma_t(1)$.

\begin{table}[b]
\begin{tabular}{lcccc}
  & $\tau_s$  & $D$ & $\tau_t$ & $z$ \\ 
  \hline
  Conserved RD  & $1.28(1)$ & $2.75(1)$ & $1.50(2)$ & $1.54(1)$ \\
  Manna & $1.28(1)$ & $2.76(1)$ & $1.48(2)$ & $1.55(1)$
\end{tabular}
\caption{Critical exponents for the conserved RD and the stochastic
  Manna models in $d=2$. Figures in parenthesis indicate the statistical 
  uncertainty in the last digit. Manna exponents from
  Refs.~\protect\cite{granada,lubeck00,nakanishi97,milshtein98}.}
\label{table2d}
\end{table}

\begin{table}[b]
\begin{tabular}{lcccc}
  & $\tau_s$  & $D$ & $\tau_t$ & $z$ \\ 
  \hline
  Conserved RD  & $1.42(1)$ & $3.36(1)$ & $1.80(2)$ & $1.77(1)$ \\
  Manna & $1.41(1)$ & $3.36(1)$ & $1.78(2)$ & $1.76(1)$
\end{tabular}
\caption{Critical exponents for the conserved RD and the stochastic
  Manna models in $d=3$. Figures in parenthesis indicate the statistical 
  uncertainty in the last digit.}
\label{table3d}
\end{table}

\begin{figure}[t]
  \centerline{\epsfig{file=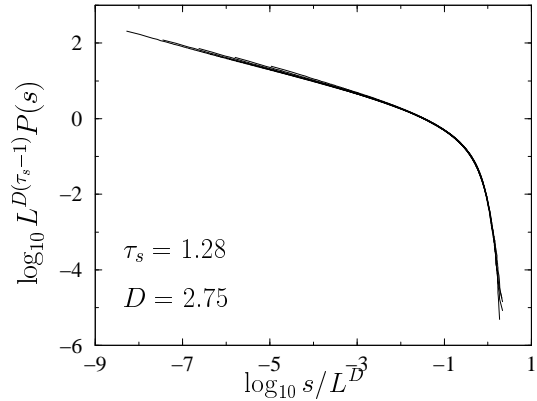, width=7cm}}
  \caption{Data collapse analysis of the integrated avalanche size
    distribution 
    for the conserved RD model in $d=2$. System sizes $L=64, 128,
    256, 512$, and $1024$.}
  \label{fig:S2d}
\end{figure}

\begin{figure}[t]
  \centerline{\epsfig{file=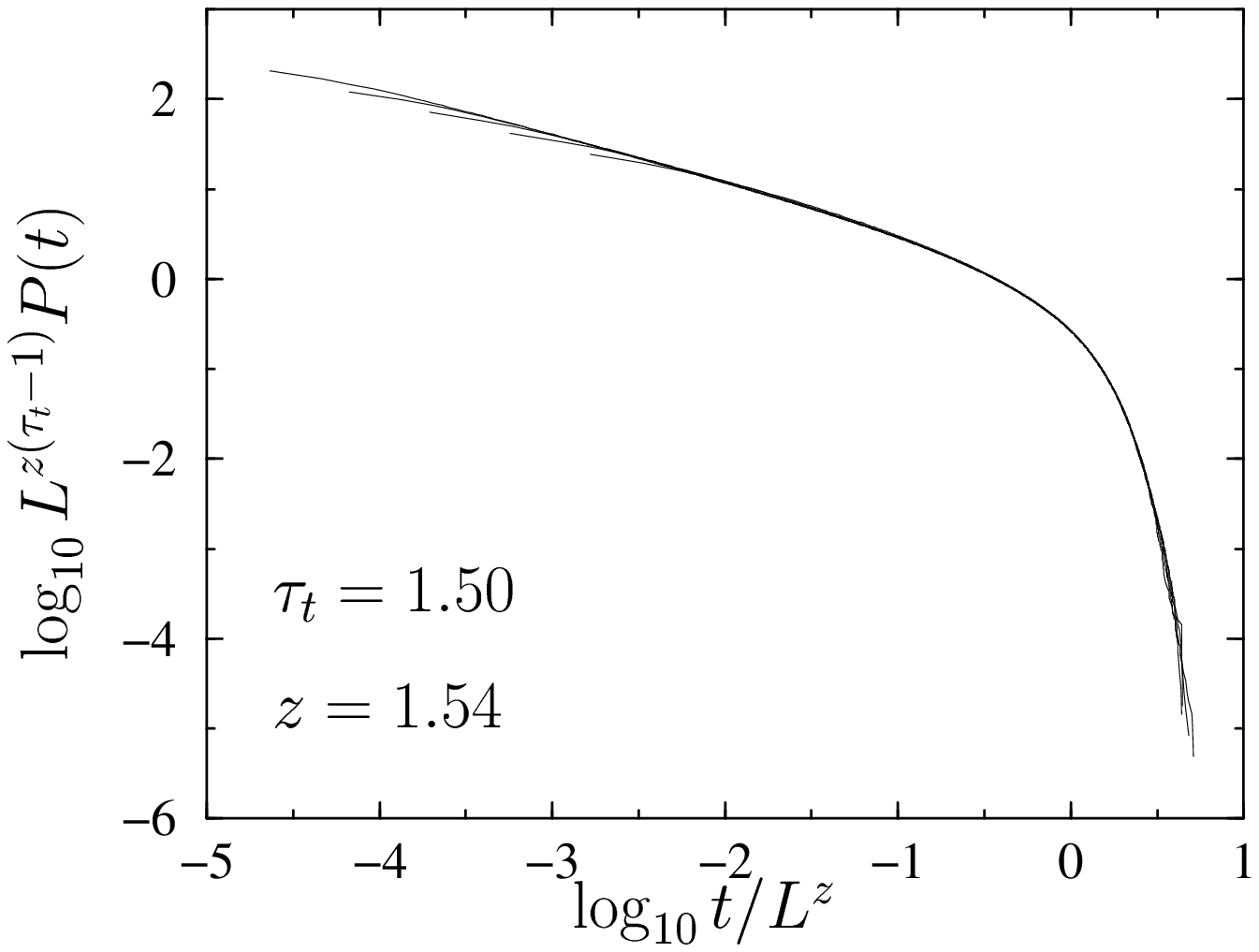, width=7cm}}
  \caption{Data collapse analysis of the integrated avalanche time
    distribution 
    for the conserved RD model in $d=2$. System sizes $L=64, 128,
    256, 512$, and $1024$.}
  \label{fig:T2d}
\end{figure}

\begin{figure}[t]
  \centerline{\epsfig{file=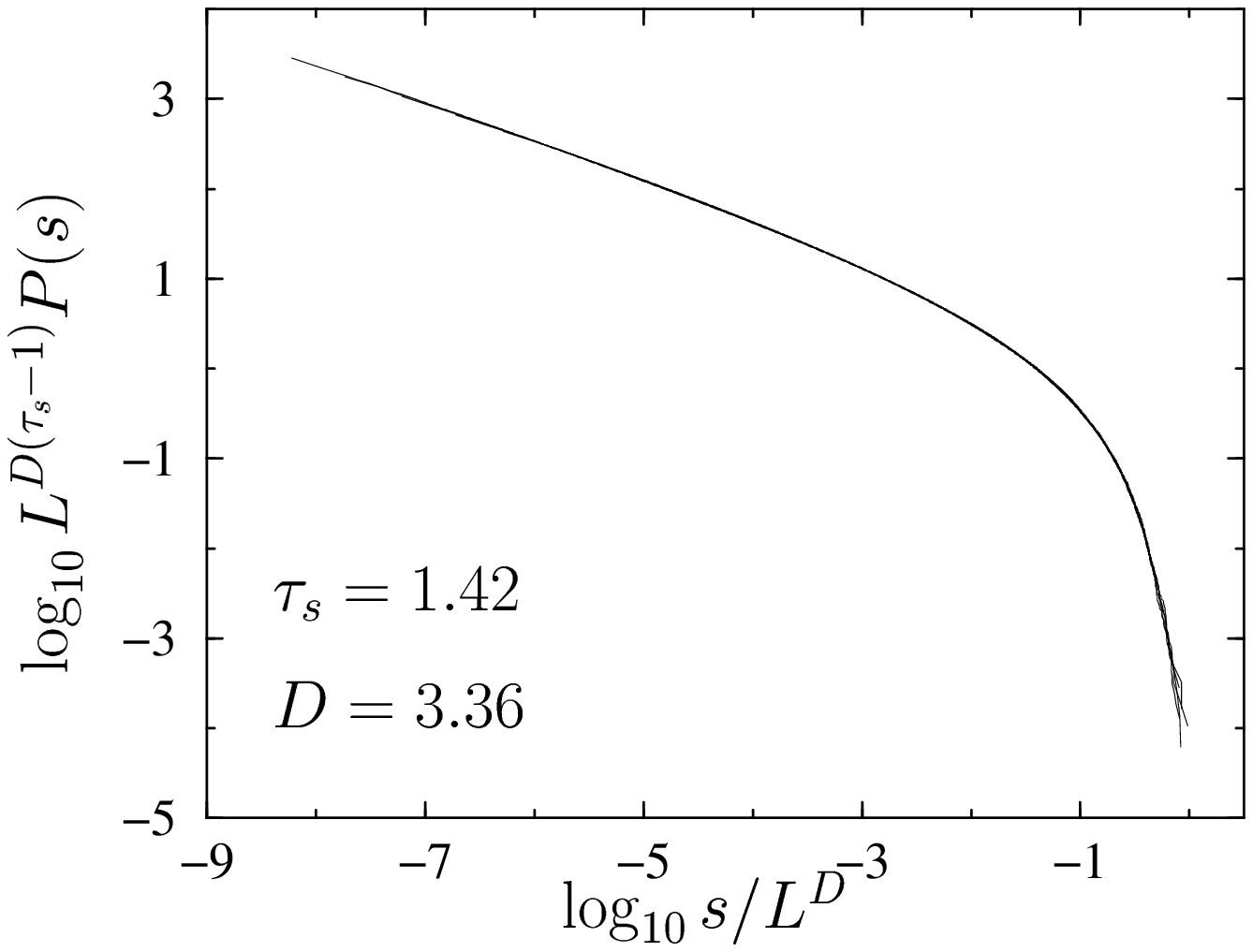, width=7cm}}
  \caption{Data collapse analysis of the integrated avalanche size
    distribution 
    for the conserved RD model in $d=3$. System sizes $L=74, 100,
    140, 200$, and $280$.}
  \label{fig:S3d}
\end{figure}

\begin{figure}[t]
  \centerline{\epsfig{file=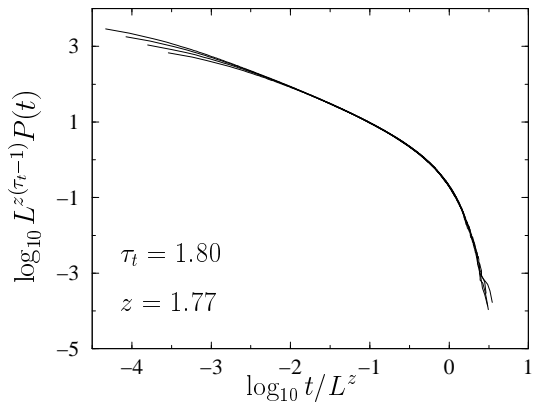, width=7cm}}
  \caption{Data collapse analysis of the integrated avalanche time
    distribution 
    for the conserved RD model in $d=3$. System sizes $L=100, 140,
    200$, and $280$.}
  \label{fig:T3d}
\end{figure}

We have computed the exponents $\tau_s$, $\tau_t$, $D$, and $z$ from
our data, using the moment analysis method. Our results are reported
in Tables~\ref{table2d} and~\ref{table3d}. The validity of these
exponents can be checked {\em a posteriori} by means of a data
collapse analysis: If the FSS hypothesis of Eqs.~\equ{eq:fssS}
and~\equ{eq:fssT} is correct, then the plots of the distributions,
under the rescaling $s\to s/L^D$ and $P(s)\to P(s)L^{D\tau_s}$, and
correspondingly $t\to t/L^z$ and $P(t)\to P(t)L^{z\tau_t}$, should
collapse onto the same universal function, for different values of
$L$.  Figs.~\ref{fig:S2d} and \ref{fig:T2d} show the data collapse for
the distributions of sizes and times in $d=2$, respectively. Analogous
plots for the case $d=3$ are presented in Figs.~\ref{fig:S3d} and
\ref{fig:T3d}. From these plots we conclude that our model fulfills
the FSS hypothesis, and is in this sense critical, being characterized
by the exponents reported in Tables~\ref{table2d} and~\ref{table3d}.

\begin{figure}[t]
  \centerline{\epsfig{file=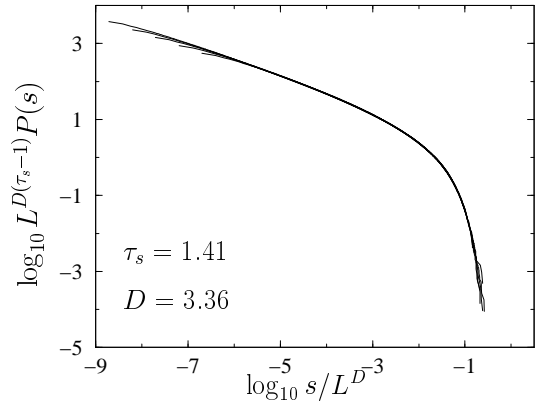, width=7cm}}
  \caption{Data collapse analysis of the integrated avalanche size
    distribution 
    for the stochastic Manna model in $d=3$. System sizes $L=100, 140,
    200, 280$, and $400$.}
  \label{fig:MannaS3d}
\end{figure}

\begin{figure}[t]
  \centerline{\epsfig{file=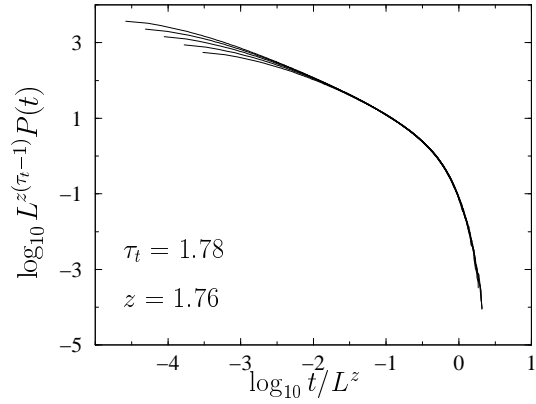, width=7cm}}
  \caption{Data collapse analysis of the integrated avalanche time
    distribution 
    for the stochastic Manna model in $d=3$. System sizes $L=100, 140,
    200, 280$, and $400$.}
  \label{fig:MannaT3d}
\end{figure}

Once we have shown that the slowly driven conserved RD model exhibits
SOC behavior, it is interesting to check whether it shares the same
universality class with any other SOC models. Given its stochastic
rules, the natural candidate for comparison is the stochastic Manna
sandpile model \cite{manna91b,dhar99}. In Table~\ref{table2d} we quote
the exponents for the Manna model in $d=2$, whose value has been
relatively well established by different sets of independent
simulations~\cite{granada,lubeck00,nakanishi97,milshtein98}.  The case
$d=3$ has not been studied so thoroughly, with the exception of the
simulations of L\"{u}beck~\cite{lubeck00}. Thus, in order to compare
our results with the conserved RD model, we have carried out
large-scale simulations of the $d=3$ stochastic Manna
model. The exponents obtained are given in Table \ref{table3d}, while
Figures~\ref{fig:MannaS3d} and~\ref{fig:MannaT3d} plot the data
collapse for sizes and times, respectively. Averages are performed
over $10^7$ non-zero avalanches, and the system sizes considered range from
$L=74$ to $L=400$ (larger than those achieved by
L\"{u}beck~\cite{lubeck00}).

From the examination of Tables~\ref{table2d} and \ref{table3d}, we
conclude that the present conserved RD model exhibits exponents fully
compatible with those of the stochastic Manna sandpile model in both
$d=2$ and $d=3$. This coincidence of exponents proves that both models
belong to the same universality class. This fact is altogether not
surprising, since both models exhibit the same basic symmetries:
isotropic diffusion of particles, stochastic conserved microscopic
rules, and presence of infinitely many---in the thermodynamic
limit---absorbing states. 
This result represents a further confirmation of the
universality claim made in
Refs.~\cite{pv00,rossi00} for this kind of systems.

In summary, we have shown how to construct a slowly-driven conserved
RD system which exhibits SOC behavior (avalanche macroscopic
dynamics). The key points are the application of the slow-driving
condition (addition of new $B$ particles) plus boundary dissipation
(diffusion of $B$ particles out of the system). The model is
characterized by critical exponents that place it in the same
universality class than the Manna stochastic sandpile model. The
related model proposed in Ref.~\cite{wij98} with $D_A \neq 0$,
however, belongs to a different universality class \cite{pv00}.  The
limit $D_A\to0$ in the theory with $D_A\neq0$ is non-analytic; any
infinitesimal amount of diffusion in the energy field renormalizes to
a finite value, and definitely changes the universality class of the
model.

This work has been supported by the European Network under Contract
No.~ERBFMRXCT980183. RP-S also acknowledges support from the grant
CICYT PB97-0693.

\newpage

\newpage

\end{document}